\RequirePackage{fix-cm}
\documentclass[onecolumn,epjc3]{svjour3}  
\smartqed  
\RequirePackage{graphicx}
\usepackage{hyperref}
\usepackage{adjustbox}
\usepackage{amsmath}
\usepackage[graphicx]{realboxes}
\usepackage{rotating}
\usepackage{float}
\usepackage{latexsym}
\usepackage{amssymb}
\usepackage{array}
\usepackage{longtable}
\usepackage{dcolumn}
\usepackage{bm}
\usepackage{tabularx}
\usepackage[font=small,labelfont=bf]{caption}
\usepackage{lscape}
\renewcommand{\figurename}{Fig.}
\RequirePackage{mathptmx} 
\journalname{}
\begin{document}

\title{ Wormholes in 4D Einstein-Gauss-Bonnet gravity with BEC Dark Matter density profile 
}


\author{Bibhash Das\thanksref{e1}
        \and
        Bikash Chandra Paul \thanksref{e2} 
}

\thankstext{e1}{e-mail: rs\_bibhash@nbu.ac.in\, (ORCID: 0000-0001-7283-6745)}
\thankstext{e2}{e-mail: bcpaul@nbu.ac.in\, (ORCID : 0000-0001-5675-5857)\, (Corresponding Author)}


\institute{Department of Physics, University of North Bengal, Darjeeling, West Bengal, India -734 013 
}

\date{Received: date / Accepted: date}

\maketitle

\begin{abstract}
The existence of Traversable Wormhole (TW) in the  4D Einstein-Gauss-Bonnet (4D-EGB) gravity is explored with phenomenological Bose-Einstein Condensates (BEC) dark matter density profile. In the framework of 4D EGB gravity, which one obtains by regularizing the higher- dimensional EGB gravity in the limit $D \to 4$ is considered to obtain a spherically symmetric TW. The Gauss-Bonnet coupling parameter ($\alpha$) in this case is rescaled to $\alpha \to \frac{\alpha}{D-4}$. Considering the energy density profile of non-relativistic BEC matter, the shape function of the WH geometry and the Null energy condition (NEC) are determined with a constant redshift function. The applicability of realistic flaring-out condition and asymptotic flatness conditions are studied here and the domain of model parameters for realistic scenario is determined. We analyze the embedding diagram of the WH obtained here with proper radial distance, volume integral quantifier, and anisotropy. We obtained WH which is stable at the throat when $\alpha = -0.0512471$ for a set of model parameters, which is estimated from sound speed measurement. The energy conditions are investigated, and it is noted that there is a range of Gauss-Bonnet coupling parameter  $\alpha \in [-4,-4.222]$, at which the energy conditions, including NEC, are obeyed at the throat. We explore NEC for other values of $\alpha$ and found that it is not satisfied. The other energy conditions are also violated. A new result is thus obtained with 4D Einstein- Gauss-Bonnet gravity with BEC Dark Matter profile. We determine the parameter space for which the WH solution exists in the proposed modified gravity model.
\end{abstract}

\section{Introduction}
\label{intro}
In theoretical astrophysics Wormhole (WH) is an interesting idea that emerges from Einstein's General theory of Relativity (GR) with a special type of matter. WH acts as a bridge connecting two distinct regions of the universe or two distinct universes. Earlier in 1916, Flamm \cite{Flamm1916} introduced a tunnel-like structure making use of Schwarzschild solution. Later, Einstein and Rosen \cite{Einstein1935} proposed a bridge-like structure that links two external spacetime regions. A hypothetical connection with a singularity-free structure between distinct regions is known as an {\it Einstein-Rosen Bridge} (ERB). Wheeler coined the name of this structure by {\it ``Wormhole"} \cite{Wheeler1957}, and subsequently shown that the bridge structure is not stable since it would collapse immediately after its formation, which leads to instability for interstellar travel \cite{Fuller1962}. In recent times, a spurt in research activities in traversable Lorentzian WH \cite{Morris1988,Visser1995} has began. However, Lorentzian WH does not possess a horizon; therefore, a two-way passage is possible. The idea of two-way passage for interstellar travel was first introduced by Morris and Thorne \cite{Morris1988}. A spherically symmetric metric that connects two asymptotically flat spacetimes at the {\it throat} of the WH is considered by them. It is shown that in such a case, exotic matter is required to keep the spacetime region open, which is linked at the throat. However, it is also found that the null energy condition (NEC) at the throat is violated. It is speculated that exotic matter in nature might exist within the context of quantum field theory \cite{Wheeler1955}. Visser $et.al.$ \cite{Visser2003} demonstrated that physically possible traversable WHs can be sustained for a specific geometry, minimizing the need for exotic matter in understanding the astrophysical object.

Recently, modified theories of gravity in the usual four dimensions and in higher dimensions have emerged to accommodate late-time acceleration, as standard GR with normal matter fails to accommodate the recent observations. Modifications in Einstein's GR permit us extra degrees of freedom in the gravitational sector that can be used to describe dark matter and dark energy required at late-time. Thus, it is interesting to explore the WH geometry in the framework of the modified theories of gravity. In the literature modified gravity, namely,  $f(R)$ gravity \cite{Lobo2009,Azizi2013,Mazharimousavi2016}, $f(R,T)$ gravity \cite{Moraes2017,Mishra2020,Chanda2021}, brane-world gravity \cite{Bronnikov2003,LaCamera2003,Parsaei2015}, Rastall gravity \cite{Javed2021,Lobo2020,Tangphati2023}, $f(Q)$ gravity \cite{Mustafa2021,Hassan2022,Banerjee2021,Hassan2023}, are considered to explore WH geometry either for a given shape function or a determined shape function arising from the stability. Traversable WH solutions are explored for two different non-linear $f(R,L_{m})$ models in the refs. \cite{Solanki2023}. WH solutions and energy conditions are studied in ref. \cite{Shamir2021}. Other related research on WH in modified gravity can be found in refs. \cite{Shaikh2018,Bahamonde2016,Jusufi2018,Jusufi2019}.

At very high enough energy, the string theory permits addition of Gauss-Bonnet term to the Einstein shell for its consistency which however is consistent in spacetime dimension more than four.  But the work of Witten \cite{Witten1981} yields that the higher dimensional theory  can be reduced to the usual four dimensions by compactification technique. 
In recent times, interest for WH geometries in higher curvature Einstein-Gauss-Bonnet (EGB) theories revived \cite{Bhawal1992,Paul1995,Zubair2023a,Zubair2023}.  EGB gravity \cite{Mehdizadeh2015,Kanti2012,Maeda2008} is the simplest natural extension of GR in higher-order curvature gravity, and it arises as quadratic terms in Lanczos-Lovelock gravity in higher dimensions \cite{Lanczos1938,Lovelock1971,Lovelock1972}. The EGB gravity is known as a ghost-free theory, i.e., it does not contain particles with negative kinetic energy terms that lead to instabilities \cite{Boulware1985}. This modified gravity theory also can be obtained in the low energy limit of the string theory \cite{Zwiebach1985}. However, in EGB gravity, the Gauss-Bonnet Lagrangian is a total derivative in 4D spacetime and does not contribute to the 4D field equations. Thus, in EGB gravity, non-trivial gravitational dynamics require $D \geq 5$. This issue was resolved by rescaling the Gauss-Bonnet coupling parameter ($\alpha$), such that $\alpha \to \frac{\alpha}{D-4}$ \cite{Glavan2020,Cognola2013}. By rescaling the coupling parameter, some interesting properties of the 4D EGB theory are noted, i.e., it bypasses the conclusions of Lovelock’s theorem and avoids the Ostrogradsky instability \cite{Ostrogradsky1850,Woodard2007}. More details about the 4D EGB gravity can be found in the ref. \cite{Fernandes2022}. The regularization process discussed in references \cite{Glavan2020,Cognola2013} is currently under scrutiny to test its validity \cite{Hennigar2020,Ai2020,Shu2020,Mahapatra2020,Arrechea2020}, and alternative regularization methods have also been suggested \cite{Hennigar2020,Lu2020,Casalino2020}. Nonetheless, the spherically symmetric 4D Black Hole solution obtained in \cite{Glavan2020,Cognola2013}, coincides with the results obtained by the alternative procedures  \cite{Hennigar2020,Lu2020,Casalino2020} and no new solutions are found. Meanwhile, 4D EGB gravity has generated substantial attention in the context of realizing the traversable WH geometry. Traversable WHs are explored in the ref. \cite{Ahmed2022} considering different shape functions, which satisfy the condition, $\mathcal{S}(r_0) = r_0$, where $r_0$ is the WH throat. Godani et al. \cite{Godani2022} investigated the stability of a thin-shell WH in 4D EGB gravity. Yukawa-Casimir WH is explored in the ref. \cite{Shweta2023}. More works on WH geometry in the context of 4D EGB gravity is found in the literature \cite{Jusufi2020,Mishra2022,Panyasiripan2024,Godani2024}. 

Bose-Einstein Condensate (BEC) is a state of matter that occurs when a system of bosons is cooled down to temperatures very close to absolute zero in a confined potential \cite{Bose1924,Einstein1925}. At these extremely low temperatures, a large fraction of bosons occupy the lowest quantum state below some critical temperature, resulting in macroscopic quantum phenomena \cite{Griffin1995}. Numerous theoretical and phenomenological studies have followed the experimental achievement of Bose-Einstein condensates (BECs) \cite{Anderson1995}. The density profile for BEC Dark Matter model can be described by the Thomas-Fermi (TF) profile \cite{Bhmer2007}. The density distribution can be derived within the Newtonian approximation by solving the Poisson equations, which is not strictly valid in strong-field regions such as WH throat. However, in the literature the BEC distribution is used as a phenomenologically motivated profile to study some essential features of BEC fluids in astrophysical objects. This is similar to  Modified Newtonian Dynamics (MOND) \cite{Milgrom1983AGalaxies,Milgrom1983AHypothesis}, where Newtonian-like equations are retained while introducing modifications that allow one to probe regimes where standard gravity might otherwise break down. Analogously, familiar non-relativistic BEC density profile can be used as an effective input to explore WHs, potentially providing the conditions to maintain a stable structure at the throat. Richarte et al. \cite{Richarte2017} constructed an asymptotically AdS/dS traversable thin-shell WH created by BECs. A quantum simulation of TW spacetime in the framework of  Bose-Einstein condensate is explored in the usual four dimension, as well as in lower dimensions (1 + 1)D \cite{Mateos2018}. Jusufi et al. \cite{Jusufi2019} obtained the necessary condition for the formation of TWs assuming BEC dark matter (BEC-DM) halo. Freitas et al. \cite{deFreitas2024} studied BECs under the effects of the non-condensate atomic cloud and showed that non-local effects in the condensed matter system define an analog model for Euclidean WH. The objective of the present work is to explore TW solutions, considering BEC density profile as a phenomenology dark matter input while allowing the 4D Einstein–Gauss–Bonnet (4D-EGB) theory to dictate how this matter behaves in a strong gravity background.

The paper is organized as follows: In Sect. \ref{sec:2}, the basic equations for 4D EGB gravity are presented. Sect. \ref{sec:3} briefly reviewed BEC as a matter of source. In Sect. \ref{sec:4}, we obtain WH solutions supported by the non-relativistic BEC matter. Here, we analyze the necessary conditions (flaring-out condition, asymptotically flatness condition) required for a traversable WH. The embedding diagram of the WH is drawn, and we determined the proper radial distance. The amount of exotic matter required to open the WH throat is estimated. In Sect. \ref{sec:5}, the energy conditions are studied with BEC matter in the 4D EGB gravity. In Sect. \ref{sec:6}, we present a brief discussion on the WH permitted in 4D EGB gravity.

\section{4D EGB gravity: Basic equations}
\label{sec:2}
The gravitational action for D-dimensional modified Einstein-Gauss-Bonnet (EGB) gravity \cite{Ghosh2020}  is given by,
    \begin{equation}
         \label{EGBaction}
         \mathcal{A} = \frac{1}{16 \pi}\int d^{D}x\,\, \sqrt{-g}\left[R+ \frac{\alpha}{D-4} \mathcal{L}_{GB} \right]  + \mathcal{A}_{m}
    \end{equation}
where, $R$ is the Ricci scalar, $\mathcal{L}_{GB}$ is the Gauss-Bonnet term, $g$ is the determinant of the metric in D-dimension, $\alpha$ is the Gauss-Bonnet coupling parameter, and  $\mathcal{A}_m$ is the action term for the matter part. The dimension of the Gauss-Bonnet coupling parameter($\alpha$) is  $[length]^2$. The Gauss-Bonnet Lagrangian ($\mathcal{L}_{GB}$) is given by,
    \begin{equation}
	\mathcal{L}_{GB} = R^{2} - 4R_{ab} \, R^{ab} + R_{abcd} \, R^{abcd}
    \end{equation}
where the indices $a$, $b$, $c$ and $d$ run from 0 to ($D-1$). Variation of the action (\ref{EGBaction}) with respect to $g_{ab}$ yields \cite{Ghosh2020}
    \begin{equation}
	\label{FE}
       \mathcal{G}_{ab} + \frac{\alpha}{D-4} \mathcal{H}_{ab} = 8 \pi \, \mathcal{T}_{ab}
    \end{equation}
where, $\mathcal{G}_{ab}$ denotes the Einstein tensor, $\mathcal{T}_{ab}$ is the total energy-momentum tensor and $\mathcal{H}_{ab}$ denotes the Lanczos tensor with the following expression:
    \begin{equation}
    	\mathcal{G}_{ab} =  R_{ab} - \frac{1}{2}R \, g_{ab},
    \end{equation}
    \begin{equation}
    	\mathcal{T}_{ab} = - \frac{2}{\sqrt{-g}}\frac{\delta(\sqrt{-g} \, \mathcal{A}_m)}{\delta g^{ab}} ,
    \end{equation}
    \begin{equation}
    	\mathcal{H}_{ab}= 2(R \, R_{ab}-2 R_{ac} \, R^{c}_{b}-2R^{cd} \, R_{abcd} + R^{cde}_{a} \, R_{bcde}) - \frac{1}{2}g_{ab} \, \mathcal{L}_{ab} .
    \end{equation} 
 In general, the Gauss-Bonnet coupling term is a total derivative in the usual 4D spacetime, resulting in no contribution in the 4D field equations. However, the coupling parameter is rescaled by $\frac{\alpha}{D-4}$, and for a maximally symmetric spacetime with curvature scalar $\mathcal{K}$ \cite{Ghosh2020}, one obtains
     \begin{equation}
        \frac{g_{ab}}{\sqrt{-g}} \frac{\delta \mathcal{L}_{GB}}{\delta g_{ab}} = \frac{\alpha (D-2)(D-3)}{2(D-1)} \mathcal{K}^2 \, \delta^a_b,
     \end{equation}
In this case a variation of the Gauss-Bonnet term is non-zero in 4-dimensional spacetime because of the rescaled coupling parameter \cite{Glavan2020}. Hereafter, we consider the gravitational unit $G_D = c^2 = 1$ in the paper.

\section{Bose–Einstein Condensate (BEC) matter}
\label{sec:3}
Bose-Einstein condensation (BEC) is a state of matter attained when diluted Bose gas or a system of bosons is cooled down to a temperature very close to absolute zero. At the extremely low temperature, a large number of bosons occupy the quantum ground state, leading to some macroscopic quantum phenomena. In this situation, the quantum state of bosons can be represented by a single particle quantum state for  $N$ interacting condensed bosons. The Hamiltonian for a many-body system of interacting bosons in an external potential $V_{ext}$ is given by \cite{Bhmer2007}
    \begin{multline}
       \hat{\mathcal{H}} = \int d\vec{r}\,  \hat{\Phi}^{+}(\vec{r}) \Bigg[ - \frac{\hbar}{2m}\nabla^2 + V_{rot}(\vec{r}) + V_{ext}(\vec{r}) \Bigg] \,   \hat{\Phi}^{-}(\vec{r}) \\
       +\, \frac{1}{2} \int d(\vec{r})\,d(\vec{r'})\, \hat{\Phi}^{+}(\vec{r})\, \hat{\Phi}^{+}(\vec{r'})\,V(\vec{r} - \vec{r'})\, \hat{\Phi}^{-}(\vec{r})\, \hat{\Phi}^{-}(\vec{r'}),
    \end{multline}
where, $\hat{\Phi}^{-}(\vec{r})$ and $\hat{\Phi}^{+}(\vec{r})$ represent the boson field operators that annihilate and create a particle at the position $\vec{r}$, respectively, and $V(\vec{r} - \vec{r'})$ represents the two-body interatomic potential. The term $V_{rot}(\vec{r})$ represents the potential corresponding to the rotation of the condensate. In the present paper, we consider  $V_{ext}(\vec{r})$ equal to the gravitational potential $V$, and neglecting the rotational potential (i.e., $V_{rot}(\vec{r}) = 0$), one obtains the Poisson equation given by
    \begin{equation}
       \nabla^2V = 4\pi\,\rho_m,
    \end{equation}
where $\rho_m$ ($= m\,\tilde{\rho}$) is the mass density of the condensate. Considering the first-order approximation and neglecting the rotational potential, the radius of the BEC can be determined, which is
\begin{equation}
    R_{BE} = \pi \sqrt{\frac{\hbar^2\,\varphi}{m^3}},
\end{equation}
where, $\varphi$ is the scattering length, which is related to the scattering cross-section of particles in the condensate. The density distribution of the BEC Dark matter is given by
    \begin{equation}
    \label{becdensity}
       \rho_{BEC} (r) = \rho_0 \frac{\sin kr}{k\,r},
    \end{equation}
where $k = \sqrt{\frac{m^3}{\hbar^2 \varphi}} = \pi/R_{BE}$, and $\rho_0$ is the central density of the condensate, i.e., $\rho_0 = \rho_{BEC}\,(0)$. We consider this BEC dark matter density profile as a phenomenological input to explore WHs within the framework of 4D-EGB gravity.

\section{Traversable Wormhole configuration in 4D EGB gravity with BEC dark metter profile}
\label{sec:4}
We consider the WH geometry in 4D Einstein-Gauss-Bonnet gravity given by a general static, spherically symmetric D-dimensional sphere,
    \begin{equation}
        \label{metric}
        ds^{2}= - e^{2\psi(r)}\,dt^{2} + \frac{dr^{2}}{\Big( 1 - \frac{\mathcal{S}(r)}{r} \Big)}+r^{2} d\Omega^{2}_{D-2},
    \end{equation}
where the potential $\psi(r)$ is the redshift parameter and $\mathcal{S}(r)$ is the shape function of the WH. $d\Omega^{2}_{D-2}$ is the metric on a unit $(D-2)$-dimensional sphere, i.e.,
\begin{equation}
    d\Omega^{2}_{D-2} = d\Theta^2_1 + \sum\limits_{i=2}^{D-2}\,\sum\limits_{j=1}^{i-1} \sin^2 (\theta_b)\,d\theta^2_a. \nonumber
\end{equation}
The radial coordinate $r$ can be extended from a minimum $r_0$ to $\infty$, where $r_0$ is the throat of the WH. For physical admissibility, we consider the following constraints on the shape function $\mathcal{S}(r)$: 
\begin{enumerate}
    \item At the throat of the wormhole, the shape function $\mathcal{S}(r_0) = r_0$  and away from the throat, i.e., $r > r_0$ it must satisfy
        \begin{equation}
            1 - \frac{\mathcal{S}(r)}{r} > 0.
        \end{equation}
    \item For traversability of the WH, the {\it flaring-out condition} is
        \begin{equation}
            \frac{\mathcal{S}(r) - r\, \mathcal{S}'(r)}{\mathcal{S}^2(r)} > 0 \,\,\,\,\,\,\,\,\text{for}\,\,\,\,\, r\geq r_0
        \end{equation}
    which yields,
        \begin{equation}
            \mathcal{S}'(r_0) < 1 \,\,\,\,\,\,\,\,\text{at}\,\,\,\,\, r = r_0,
        \end{equation}
    \item {\it Asymptotic flatness} of the spacetime geometry is associated with when
        \begin{equation}
            \frac{\mathcal{S}(r)}{r} \rightarrow 0 \,\,\,\,\,\,\,\, \text{as}\,\,\,\,\, r \rightarrow \infty
        \end{equation}
\end{enumerate}
We consider here a redshift function, $\psi(r)$, which is finite and non-zero everywhere to avoid event horizon.

For  anisotropic fluid distribution, the energy-momentum tensor  is given by
    \begin{equation}
        \label{energytensor}
        \mathcal{T}^a_b = (\rho + P_r)u_a u^b + P_{\perp} \delta^b_a + (P_r - P_{\perp})v_a v^b,
\end{equation}
where, $\rho$, $P_{r}$, and $P_{\perp}$ is the energy density, radial pressure, and transverse pressure, respectively, for the anisotropic fluid. $u^{a}$ is the d-velocity vector and $v^{a}$ represents the radially directed unit velocity space-like vector normal to $u^{a}$. For the limits $D \to 4$, the metric given by Eq. (\ref{metric}) and the energy-momentum tensor in Eq. (\ref{energytensor}), the field equations in Eq. (\ref{FE}) yield the following:
    \begin{equation}
        \label{egbdensity}
        8\pi\,\rho(r) = \frac{\alpha\,\mathcal{S}(r)}{r^6} \left( 2r\,\mathcal{S}'(r) - 3b(r) \right) + \frac{\mathcal{S}'(r)}{r^3},
    \end{equation}
    \begin{equation}
        \label{egbpressure1}
        8\pi\, P_r(r) = \frac{\alpha\,\mathcal{S}(r)}{r^6} \left( 4\psi'\,r(r-\mathcal{S}(r)) + \mathcal{S}(r) \right) + \frac{2\psi'\,(r-\mathcal{S}(r))}{r^2} - \frac{\mathcal{S}(r)}{r^3},
    \end{equation}
    \begin{multline}
        \label{egbpressure2}
        8\pi\, P_{\perp}(r) = \Big( 1 - \frac{\mathcal{S}(r)}{r} \Big)\,\Bigg( (\psi'' + \psi'^2)\Big( 1 + \frac{4\alpha\,\mathcal{S}(r)}{r^3} \Big) + \frac{1}{r} \Big( \psi' - \frac{r\,\mathcal{S}'(r)-\mathcal{S}(r)}{2r(r-\mathcal{S}(r))} \Big) \\
        \Big( 1 - \frac{2\alpha\,\mathcal{S}(r)}{r^3} \Big) - \frac{r\,\mathcal{S}'(r)-\mathcal{S}(r)}{2r(r-\mathcal{S}(r))}\,\psi'\, \Big( 1-\frac{8\alpha}{r^2} + \frac{12\alpha\,\mathcal{S}(r)}{r^3} \Big) \Bigg) - \frac{2\alpha\,\mathcal{S}^2(r)}{r^6}
    \end{multline}
where prime ($'$) denotes differentiation with respect to $r$.

    \begin{figure}[t]
        \centering
        \begin{tabular}{cc}
            \includegraphics[width=0.4\linewidth]{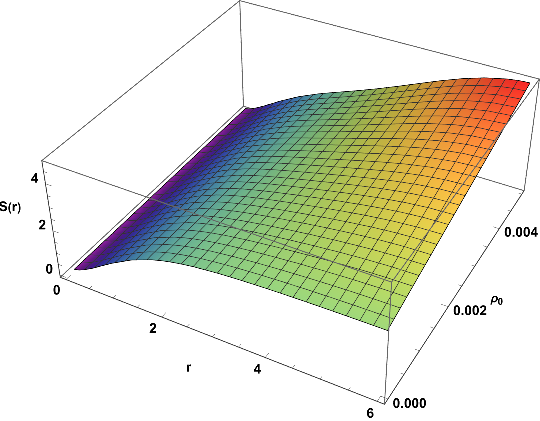}   &
            \includegraphics[width=0.4\linewidth]{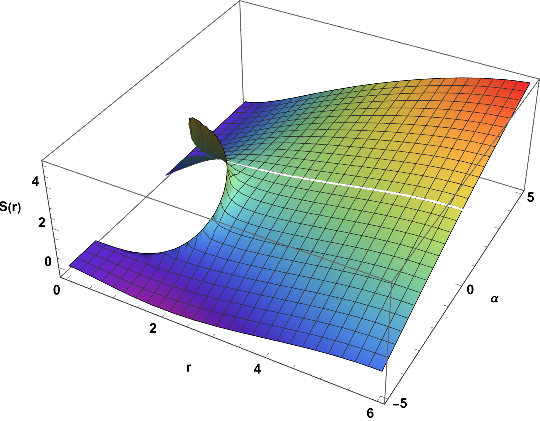} 
        \end{tabular}
             \includegraphics[width=0.4\linewidth]{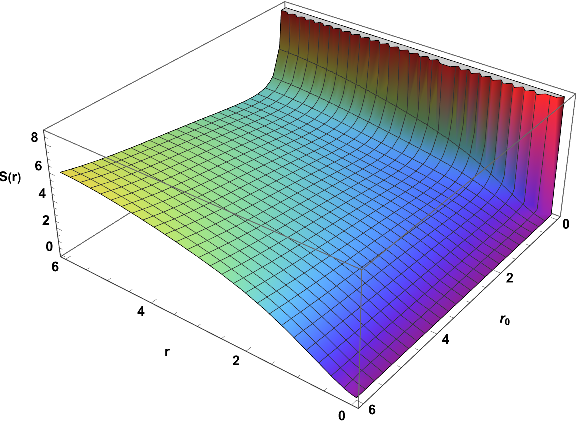}
        \caption{Shape function for different model parameters: $(R_{BE},r_0,\alpha) = (6,2,1)$(Top-left plot),  $(R_{BE},r_0,\rho_0) = (6,2,0.0025)$(Top-right plot), \& $(R_{BE},\rho_0,\alpha) = (6,0.0025,1)$(Bottom plot)  }
        \label{fig:shapefunc}
    \end{figure}

\subsection{{\bf WH solutions with BEC matter}}
\label{sec:41}
In this section, we consider the energy density of the wormhole matter, which is given by the density profile of BEC matter. On integrating Eq. (\ref{egbdensity}) using the BEC density profile given by Eq. (\ref{becdensity}), we got
    \begin{equation}
        \mathcal{S}(r) = \frac{r^3}{2\pi \alpha} \Bigg[ \frac{\sqrt{\alpha}}{r^2}\, \sqrt{r \left( \frac{\pi^2}{\alpha}(r^3 + 4\alpha^2\,C_1) + 32 R_{BE}^2\,\rho_0 \left( R_{BE} \sin \left(\frac{\pi\,r}{R_{BE}} \right) - \pi\,r \cos \left(\frac{\pi\,r}{R_{BE}} \right) \right) \right)}\, - \,\pi \Bigg],
    \end{equation}

    \begin{figure}[th]
        \centering
        \begin{tabular}{cc}
            \includegraphics[width=0.4\linewidth]{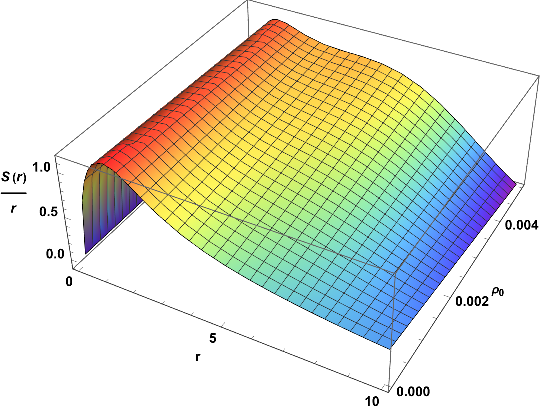}   &
            \includegraphics[width=0.4\linewidth]{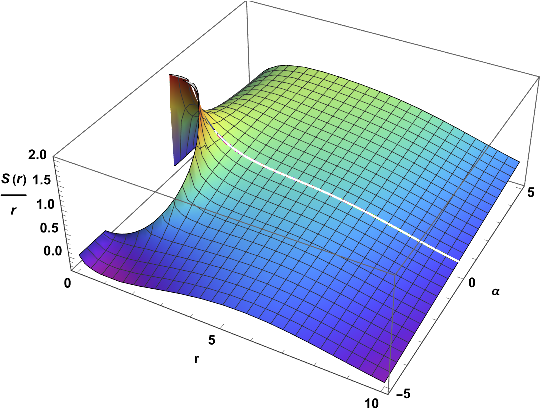} 
        \end{tabular}
             \includegraphics[width=0.4\linewidth]{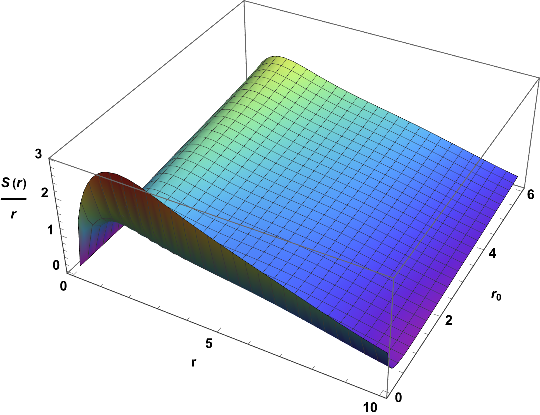}
        \caption{Asymptotic flatness condition for different model parameters: $(R_{BE},r_0,\alpha) = (6,2,1)$(Top-left plot),  $(R_{BE},r_0,\rho_0) = (6,2,0.0025)$(Top-right plot), \& $(R_{BE},\rho_0,\alpha) = (6,0.0025,1)$(Bottom plot) }
        \label{fig:bbyr}
    \end{figure} 
where $C_1$ is an integration constant and $R_{BEC}$ is the BEC radius where $\rho = 0$. Using the first condition $\mathcal{S}(r_0) = r_0$ at the throat, we determine the integration constant, $C_1$, which becomes
    \begin{equation}
        C_1 = \frac{1}{\alpha} \Bigg[  r_0  +\frac{\alpha }{r_0} + \frac{8 \rho_0\, R_{BE}^2 }{\pi ^2} \left(\pi  r_0 \cos \left(\frac{\pi  r_0}{R_{BE}}\right)-R_{BE} \sin \left(\frac{\pi r_0}{R_{BE}}\right)\right)\Bigg],
    \end{equation}
   
The shape function can now be expressed as
    \begin{multline}
    \label{shapefunction}
        \mathcal{S}(r) = \frac{r^3}{2\pi \alpha} \Bigg[ \frac{\sqrt{\alpha}}{r^2}\, \Bigg( r \Big( \pi^2 (4r_0+\frac{r^3}{\alpha} + \frac{4\alpha}{r_0}) + 32\,r\, R_{BE}^2\,\rho_0 \Big( R_{BE} \sin \left(\frac{\pi\,r}{R_{BE}} \right) - R_{BE} \sin \left(\frac{\pi\,r_0}{R_{BE}} \right)\\ 
        +\pi\,r_0 \cos \left(\frac{\pi\,r_0}{R_{BE}} \right) - \pi\,r \cos \left(\frac{\pi\,r}{R_{BE}} \right) \Big) \Big) \Bigg)^{1/2}\, - \,\pi \Bigg],
    \end{multline}
Now one gets $\frac{\mathcal{S}(r)}{r}$ at the limit $r \to R_{BE}$,
    \begin{multline}
        \lim_{r\to R_{BE}} \frac{\mathcal{S}(r)}{r} = \frac{R_{BE}^2}{2\pi \alpha} \Bigg[ \frac{\sqrt{\alpha}}{R_{BE}^2}\, \Bigg(  \pi^2\,R_{BE} (4r_0+\frac{R_{BE}^3}{\alpha} + \frac{4\alpha}{r_0}) + 32\, R_{BE}^3\,\rho_0 \Big( 
        \pi\,r_0 \cos \left(\frac{\pi\,r_0}{R_{BE}} \right)\\
     + R_{BE} \Big(\pi- \sin \left(\frac{\pi\,r_0}{R_{BE}} \right) \Big) \Big) \Bigg)^{1/2}\, - \,\pi \Bigg].
    \end{multline}

    \begin{figure}[t]
        \centering
        \begin{tabular}{cc}
            \includegraphics[width=0.4\linewidth]{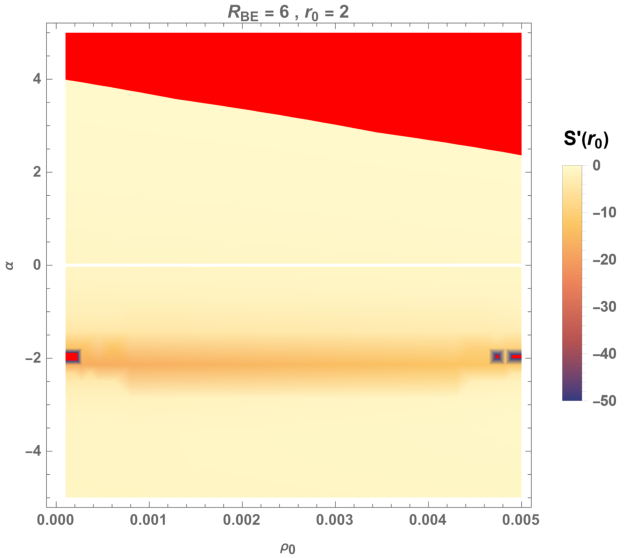}   &
              \includegraphics[width=0.4\linewidth]{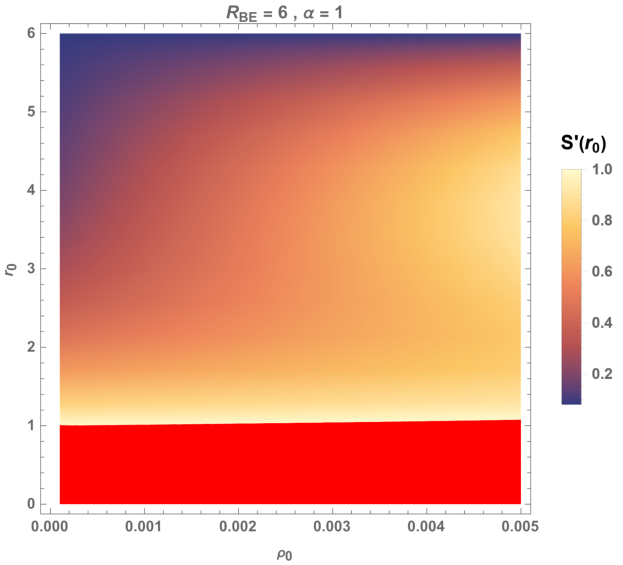} 
        \end{tabular}
             \includegraphics[width=0.4\linewidth]{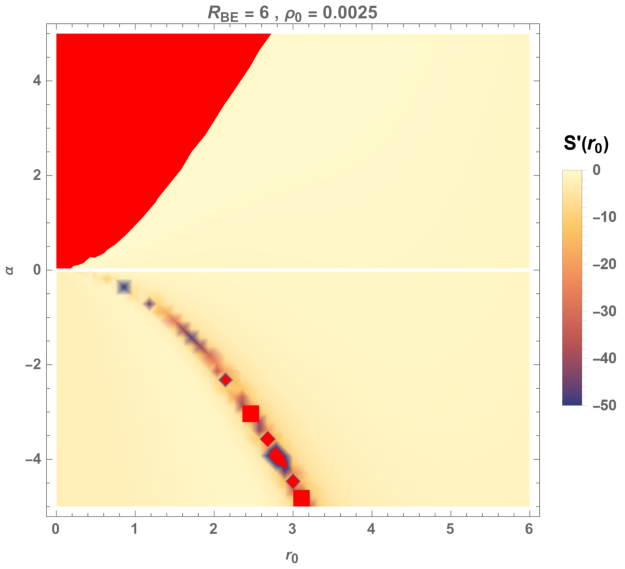}
        \caption{flaring-out condition for different model parameters: $(R_{BE},r_0) = (6,2)$(Top-left plot),  $(R_{BE},\alpha) = (6,1)$(Top-right plot), \& $(R_{BE},\rho_0) = (6,0.0025)$(Bottom plot). Red area suggests: $\mathcal{S}'(r) \geq 1$, which is unacceptable.}
        \label{fig:bprime}
    \end{figure}

The solution obtained above is not exactly asymptotically flat. For the traversability criterion, the flaring-out condition is imposed, and we determine the shape function, which is given by
    \begin{equation}
        \label{flaringcond}
        \mathcal{S}'(r) = \frac{1}{2\alpha (r_0^2 + 2\alpha)} \Bigg[ 3\,r_0^4 + 6\,r_0^2\,\alpha + 6\,\alpha^2 - 16\,R_{BE}\,r_0^3\,\alpha\,\rho_0\,\sin \left( \frac{\pi\,r_0}{R_{BE}} \right) - 3r_0^2 \,(r_0^2 + 2\alpha) \Bigg]\, < \, 1.
    \end{equation}
    
For simplicity, no tidal force exists when we choose the redshift function as $\psi = 1$. Consequently, the field equations given by Eqs. (\ref{egbdensity}) - (\ref{egbpressure2}) yield
    \begin{equation}
        \label{fe1}
        \rho(r) = \rho_0 \frac{\sin kr}{k\,r},
    \end{equation}
    \begin{equation}
        \label{fe2}
        P_r(r) = \frac{1}{32 \pi^3\,\alpha} \Big( \pi - \epsilon_1 \Big) \Big( \pi + 2\pi r^2 - \epsilon_1 \Big)
    \end{equation}
    \begin{multline}
        \label{fe3}
        P_{\perp}(r) = -\frac{1}{16\pi^3} \Bigg[ \frac{1}{\alpha} (\pi - A(r))^2 + \frac{\sqrt{\alpha}}{r_0\,r^2\,\epsilon_1}\Bigg( \alpha\,r_0 \Big( \frac{\pi r^3\,\epsilon_1}{\alpha} - 8\rho_0\,R_{BE} (\pi^2\,r^2 + R_{BE}^2) \sin \left( \frac{\pi\,r}{R_{BE}}\right) \\
        + 8\rho_0\,R_{BE}^3 \sin \left( \frac{\pi\,r_0}{R_{BE}}\right) + 8\pi\,\rho_0\,r\,R_{BE}^2 \cos \left( \frac{\pi\,r}{R_{BE}}\right) - 8\pi\,\rho_0\,r_0\,R_{BE}^2 \cos \left( \frac{\pi\,r_0}{R_{BE}}\right) \Big) - \pi^2 (r_0\,r^3 + \alpha(\alpha + r_0^2)) \Bigg) \Bigg]
    \end{multline}
where we define, $A(r) = \frac{\sqrt{\alpha}}{r^2}\, \Bigg[ r \Big( \pi^2 (4r_0+\frac{r^3}{\alpha} + \frac{4\alpha}{r_0}) + 32r\, R_{BE}^2\,\rho_0 \Big( R_{BE} \sin \left(\frac{\pi\,r}{R_{BE}} \right) - R_{BE} \sin \left(\frac{\pi\,r_0}{R_{BE}} \right)  +\pi\,r_0 \cos \left(\frac{\pi\,r}{R_{BE}} \right) - \pi\,r \cos \left(\frac{\pi\,r}{R_{BE}} \right) \Big) \Big) \Bigg]^{1/2}$.\\

In Fig. \ref{fig:shapefunc}, we plot the shape function for a domain of model parameters: $(R_{BE},\,r_0,\,\alpha) \equiv (6,2,1)$ and $\rho_0 \in [0.0001,0.005]$ (top-left); $(R_{BE},\,r_0,\,\rho_0) \equiv (6,2,0.0025)$ and $\alpha \in [-5,5]$ (top-right); $(R_{BE},\,\rho_0,\,\alpha) \equiv (6,0.0025,1)$ and $r_0 \in [0,6]$ (bottom). In Fig. \ref{fig:bbyr}, we plot the asymptotic flatness condition for the WH for the above set of model parameters. In Fig. \ref{fig:bprime}, we plot the flaring-out condition to examine the traversability of the WH. We note that for a domain of the model parameters ($R_{BE},\, r_0,\, \rho_0,\, \alpha$) discussed above, the flaring-out condition ($\mathcal{S}'(r) < 1$) is satisfied. But for other combinations of the model parameters, the flaring-out condition is not satisfied, represented by the red region in Fig. \ref{fig:bprime}.

\subsection{{\bf Embedding diagram of WH}}
\label{sec:42}
\begin{figure}[t]
        \centering
        \begin{tabular}{cc}
        \includegraphics[width=0.4\linewidth]{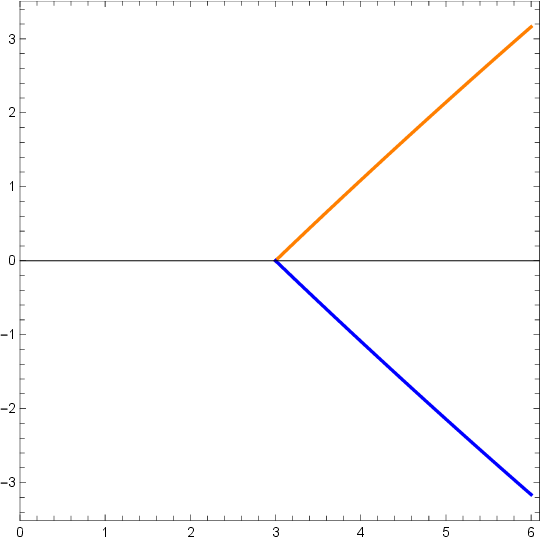}   &
            \includegraphics[width=0.4\linewidth]{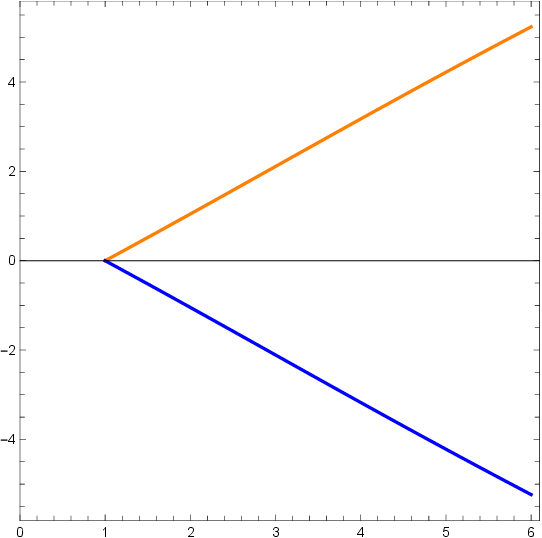} \\
            \includegraphics[width=0.4\linewidth]{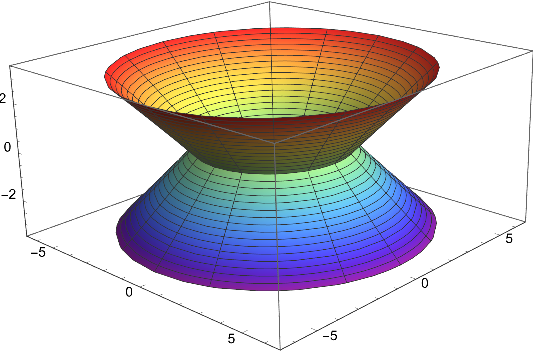}   &
            \includegraphics[width=0.4\linewidth]{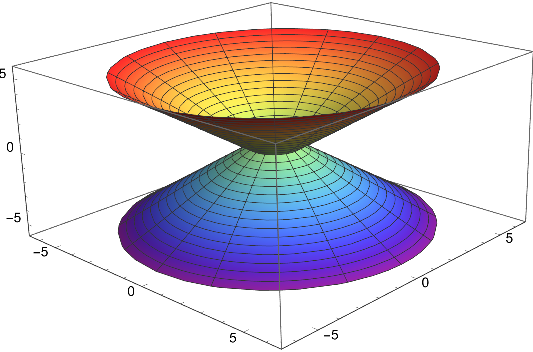} 
        \end{tabular}
        \caption{Embedding diagram of the WH for different model parameters: $(R_{BE},r_0,\alpha,\rho_0) = (6,3,1,0.0025)$(top-left and bottom-left plots), \&  $(R_{BE},r_0,\alpha,\rho_0) = (6,1,-1,0.0025)$(top-right and bottom-right plots)}
        \label{fig:WHvisual}
    \end{figure}

We analyze the embedding diagram in this subsection to understand the topology of the WH. Consider an equatorial slice at $\theta = \pi/2$ at a fixed moment (i.e., $t = const.$) of a 2D hypersurface $\zeta$. For $t = const.$, the metric (\ref{metric}) can be expressed as
    \begin{equation}
        \label{embeddedmetric1}
        ds_{\zeta}^2 = \frac{dr^{2}}{\Big( 1 - \frac{\mathcal{S}(r)}{r} \Big)} + r^{2} d\psi^{2},
    \end{equation}
The reduced metric in Eq. (\ref{embeddedmetric1}) can be embedded in a 3D cylindrically symmetric Euclidean spacetime, which is
    \begin{equation}
        ds_{\zeta}^2 = dz(r)^2 + dr^2 + r^{2} d\psi^{2},
    \end{equation}
Once again, the above equation can be expressed as
    \begin{equation}
        \label{embeddedmetric2}
         ds_{\zeta}^2 = \Bigg( 1 + \Big( \frac{dz(r)}{dr} \Big)^2 \Bigg) dr^2 + r^{2} d\psi^{2},
    \end{equation}
Matching Eqs. (\ref{embeddedmetric1}) and (\ref{embeddedmetric2}), yields
    \begin{equation}
        \frac{dz(r)}{dr} = \pm \sqrt{\frac{\mathcal{S}(r)}{r - \mathcal{S}(r)}}.
    \end{equation}
The embedded surface diagrams of the WH are plotted in the top part of Fig. \ref{fig:WHvisual}, whereas the full visualization of the WH can be found at the bottom part for different
model parameters. 

\subsection{{\bf Proper radial distance}}
\label{sec:43}
\begin{figure}[t]
        \centering
        \begin{tabular}{cc}
            \includegraphics[width=0.4\linewidth]{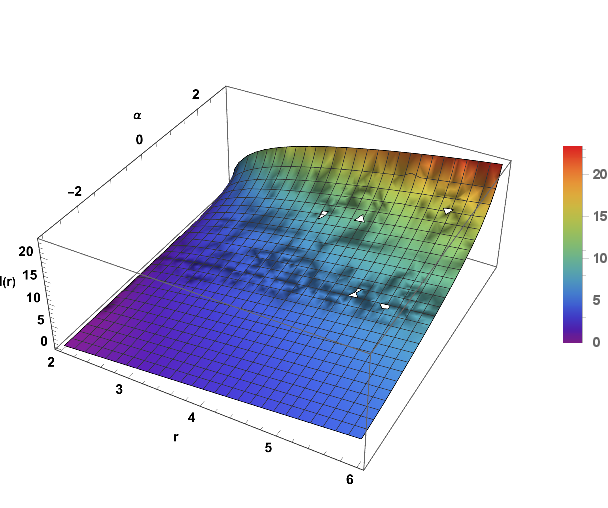}   &
            \includegraphics[width=0.4\linewidth]{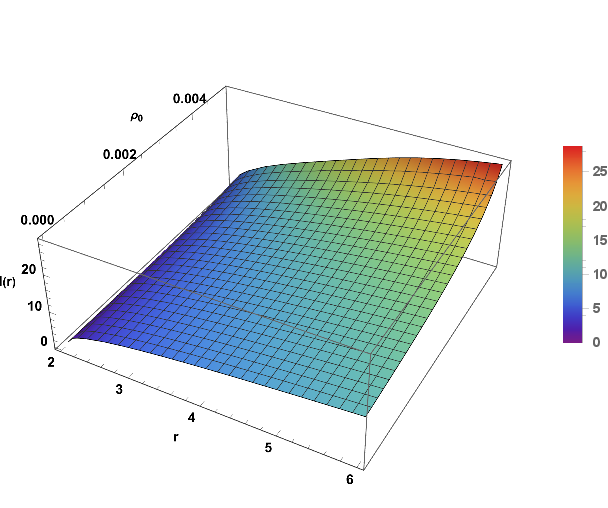} 
        \end{tabular}
        \caption{Proper radial distance of the WH for different model parameters: $(R_{BE},r_0,\alpha) = (6,2,2)$(left plot), \& $(R_{BE},r_0,\rho_0) = (6,2,0.0025)$(right plot)  }
        \label{fig:radialdistance}
    \end{figure}

Another criterion of a realistic WH is the existence of a finite proper radial distance ($l(r)$). The  proper radial distance is defined as:
    \begin{equation}
        l(r) = \pm \int^{r}_{r_0} \frac{dr}{\sqrt{1 - \frac{\mathcal{S}(r)}{r}}}.
    \end{equation}
Here, the "$\pm$"  indicates the upper and lower parts of the WH, which are connected by a throat. It is found that the proper radial distance decreases for the upper universe from $l = +\infty$ up to the throat and then from $l = 0 $ to $- \infty$ in the lower region. In Fig. \ref{fig:radialdistance} we plot the proper radial distance for a domain of model parameters:  $r_0 = 2$, $\alpha = 2$, $R_{BE} = 6$, and taking the range of values $\rho_0 \in [-3,3]$ (left plot) and $r_0 = 2$, $\rho_0 = 0.0025$, $R_{BE} = 6$, and $\alpha \in [0.0001,0.005]$ (right plot). We note that the proper radial distance decreases from the upper region towards the throat of the WH, which is physically viable.

    \begin{figure}[ht]
        \centering
        \begin{tabular}{cc}
            \includegraphics[width=0.4\linewidth]{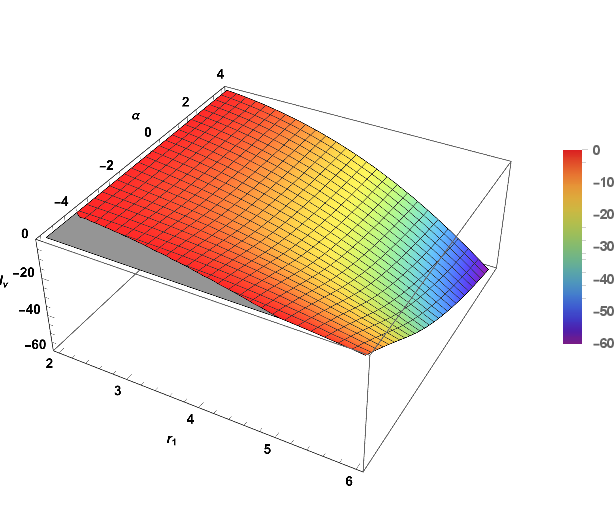}   &
            \includegraphics[width=0.4\linewidth]{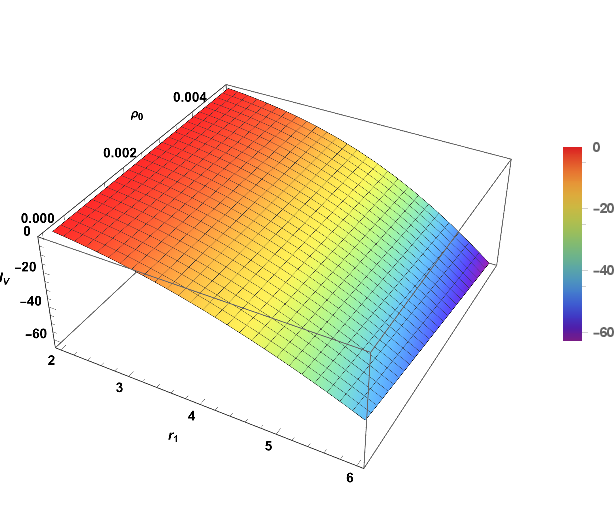} 
        \end{tabular}
        \caption{VIQ profile of the WH for different model parameters: $(R_{BE},r_0,\rho_0) = (6,2,0.0025)$(left plot), \& $(R_{BE},r_0,\alpha) = (6,2,2)$(right plot)  }
        \label{fig:viq}
    \end{figure}
    
\subsection{{\bf Measure of exotic matter}}
\label{sec:44}
The total amount of exotic matter required to stabilize a WH is calculated using the volume integral quantifier (VIQ). VIQ quantifies the average amount of exotic matter that is present in the spacetime, which, however, violates the NEC. We define VIQ by the integral,
    \begin{equation}
        I_V = \oint (\rho + P_r)\, dV,
    \end{equation}
where $dV = r^2\,\sin \theta\, dr\,d\theta \, d\psi$. Since, $\oint dV = 2 \int_{r_0}^{\infty} dV = 8\pi \int_{r_0}^{\infty} r^2\, dr $, we have
    \begin{equation}
        I_V = 8\pi \int_{r_0}^{\infty} (\rho + P_r) r^2\, dr. 
    \end{equation}
Since, the WH extends from the throat $r_0$ to a certain radius $r_1$ with $r_1 \geq r_0$, the VIQ can be expressed as
    \begin{equation}
        \label{viq}
        I_V = 8\pi \int_{r_0}^{r_1} (\rho + P_r) r^2\, dr.
    \end{equation}
We plot the radial variation of the VIQ in Fig. \ref{fig:viq} for a domain of model parameters: $r_0 = 2$, $\rho_0 = 0.0025$, $R_{BE} = 6$, and $\alpha \in [0.0001,0.005]$ (left plot) and $r_0 = 2$, $\alpha = 2$, $R_{BE} = 6$, and $\rho_0 \in [-3,3]$ (right plot). It is noted that $I_V \to 0$, as $r_1 \to r_0$. We note that a small fraction of exotic matter is required for the stable configuration of a traversable WH. We also note from `gray region' in the left plot of Fig. \ref{fig:viq} that VIQ is positive ($I_V > 0$) for $\alpha \leq -4 $, which indicates that for this range of the Gauss-Bonnet coupling parameter, the NEC does not violate.

\subsection{{\bf Anisotropy Analysis}}
\label{sec:45}
In this section, we analyze the pressure anisotropy \cite{Das2022,Das2024} in a WH supported by BEC matter within the framework of 4D EGB gravity. The Bose-Einstein Condensates are isotropic fluids and in the Thomas-Fermi (TF) approximation, it satisfy a polytropic equation of state ($P = \kappa \rho^{1 + \frac{1}{n}}$), with polytropic index $n=1$, in which the pressure is proportional to the square of the density. In the present work, the radial and transverse pressure are derived from the field equations in (\ref{FE}) using the BEC density profile as source. This indicates that the pressure obtain here, already account for the modifications introduced by the higher curvature terms in 4D-EGB gravity. Consequently, the pressure deviate from the simple Newtonian polytropic form and there is anisotropy in the system. Similar deviation from the usual features of BECs in GR, where transverse pressure is not equal to the radial pressure, have been observed in the Ref. \cite{Jusufi2019}. The anisotropy factor is given by
    \begin{equation}
        \Delta = P_{\perp} - P_{r}.
    \end{equation}
    
     \begin{figure}[t]
        \centering
        \begin{tabular}{cc}
            \includegraphics[width=0.38\linewidth]{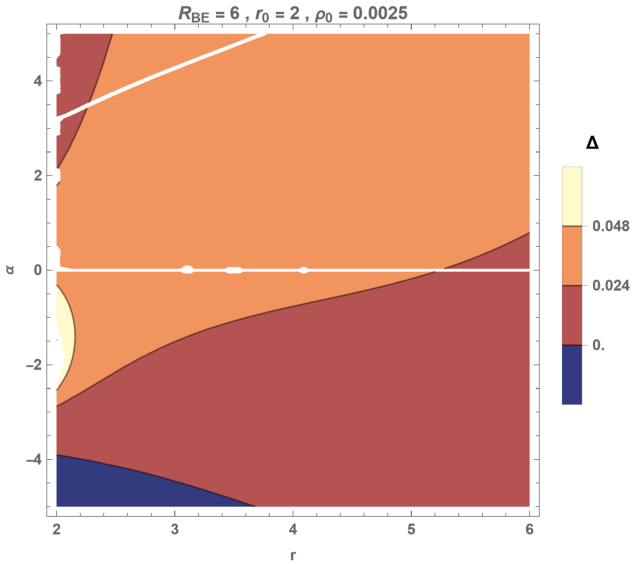}   &
              \includegraphics[width=0.4\linewidth]{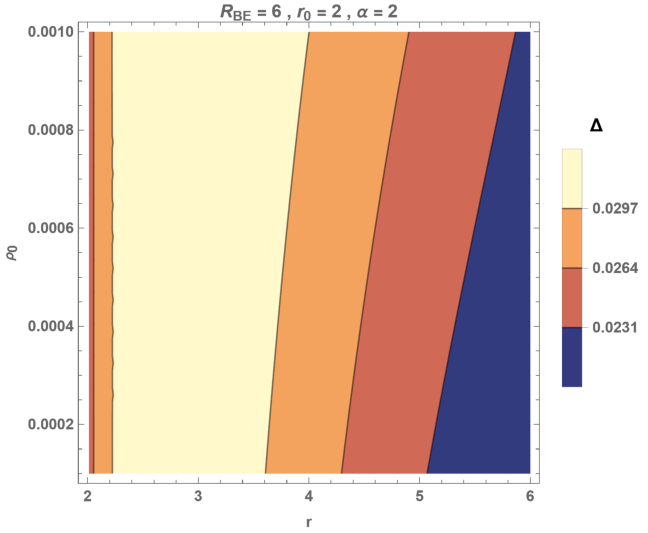} 
        \end{tabular}
        \caption{Anisotropy profile for different model parameters: $(R_{BE},r_0,\rho_0) = (6,2,0.0025)$(left plot), \& $(R_{BE},\rho_0,\alpha) = (6,0.0025,2)$(right plot)}
        \label{fig:anisotropy}
    \end{figure}
In Fig. \ref{fig:anisotropy}, it is evident that the anisotropy factor may be positive as well as negative depending upon the model parameters. A positive anisotropy ($\Delta > 0$) means a repulsive outward force that arises from anisotropy, and a negative anisotropy factor ($\Delta < 0$) signifies an attractive inward force. For the Gauss-Bonnet coupling parameters satisfying the inequality $\alpha \leq -4$, lead to a negative anisotropy factor yielding the inward attractive force.  But for $\alpha > -4$, one obtains an outward repulsive force that stabilizes the wormhole geometry.

\subsection{{\bf Vanishing sound speed}}
\label{sec:46}
\begin{figure}[t]
        \centering
        \includegraphics{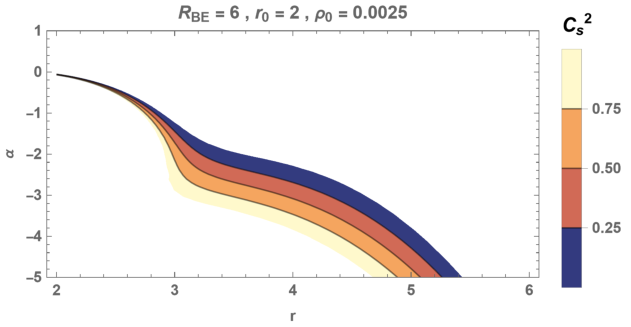}
        \caption{Contour plot of the adiabatic sound speed ($C_{s}^{2}$) for model parameters: $(R_{BE},r_0,\rho_0) = (6,2,0.0025)$ }
        \label{fig:stability}
    \end{figure}

The stability analysis of a traversable WH can be investigated by measuring the adiabatic sound speed \cite{Capozziello2021}, which is given by
    \begin{equation}
        C_{s}^{2} = \frac{\delta P}{\delta \rho},
    \end{equation}
where pressure $P$ is, $P = \frac{P_r + 2P_{\perp}}{3}$. The stability condition requires, 
    \begin{equation}
        C_{s}^{2} = 0
    \end{equation}
at the WH throat. Using Eqs. (\ref{fe2}) and (\ref{fe3}), we found that the WH model is stable at the throat of the WH for the Gauss-Bonnet coupling parameter $\alpha = -0.0512471$ for the model parameters $(R_{BE},r_0,\rho_0) = (6,2,0.0025)$. We also note that the WH model is causal ($0 \leq C_s^2 < 1$) when $\alpha \in [-0.0512471,-0.0770328]$. In Fig. \ref{fig:stability}, we plot a contour of the causal limit of $C_{s}^{2}$ for the model parameters discussed above.

\section{Energy Conditions}
\label{sec:5}
We investigate the energy conditions (ECs) for anisotropic matter distribution in WHs permitted here. The energy conditions are
    \begin{itemize}
        \item Null Energy Condition (NEC): $\rho + P_r \geq 0$, \,$\rho + P_{\perp} \geq 0$
        \item Weak Energy Condition (WEC): $\rho \geq 0$,\,$\rho + P_r \geq 0$, \,$\rho + P_{\perp} \geq 0$
        \item Strong Energy Conditions (SEC): $\rho + P_r \geq 0$, \,$\rho + P_{\perp} \geq 0$,\, $\rho + P_r + 2 P_{\perp} \geq 0$
        \item Dominant Energy Conditions (DEC): $\rho - |P_r| \geq 0$,\,$\rho - |P_{\perp}| \geq 0$
    \end{itemize}
    
    \begin{figure}[t]
         \centering
        \begin{tabular}{cc}
            \includegraphics[width=0.5\linewidth]{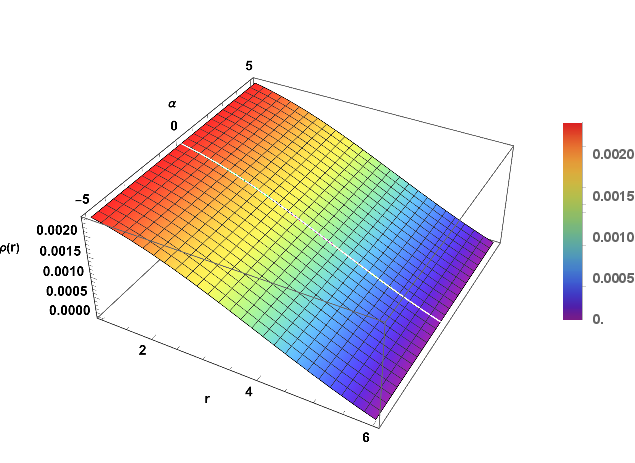}   &
            \includegraphics[width=0.5\linewidth]{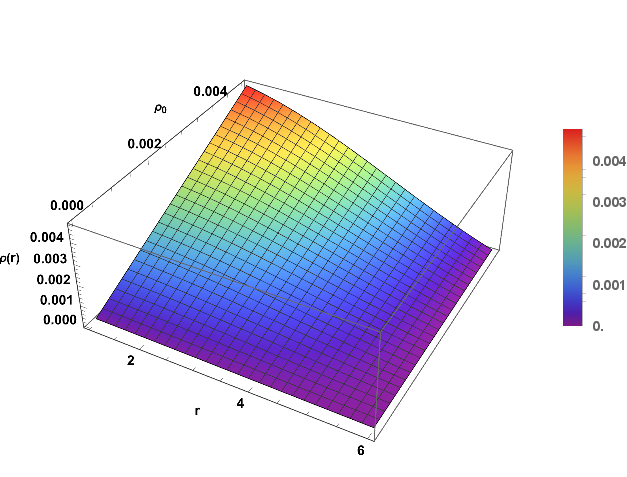} 
        \end{tabular}
        \caption{Radial variation of $(\rho(r)$ for different model parameters: $(R_{BE},r_0,\rho_0) = (6,2,0.0025)$(left plot), \& $(R_{BE},r_0,\alpha) = (6,2,2)$(right plot). }
        \label{fig:ec1}
    \end{figure}
    
     \begin{figure}[t]
        \centering
        \begin{tabular}{cc}
            \includegraphics[width=0.4\linewidth]{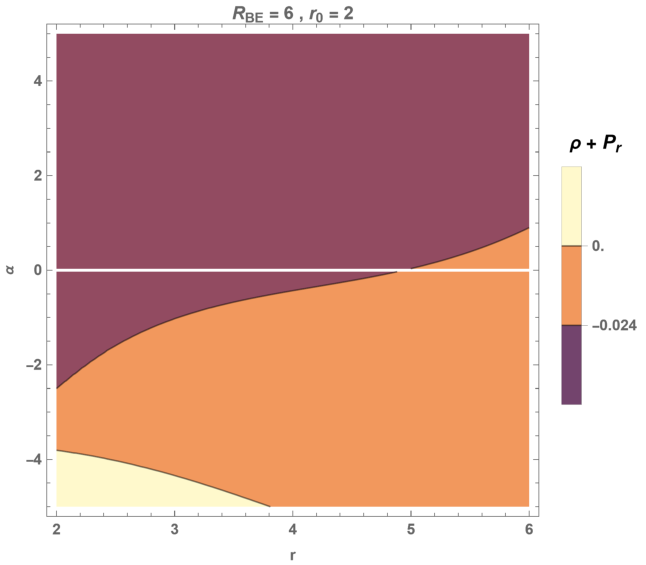}   &
            \includegraphics[width=0.4\linewidth]{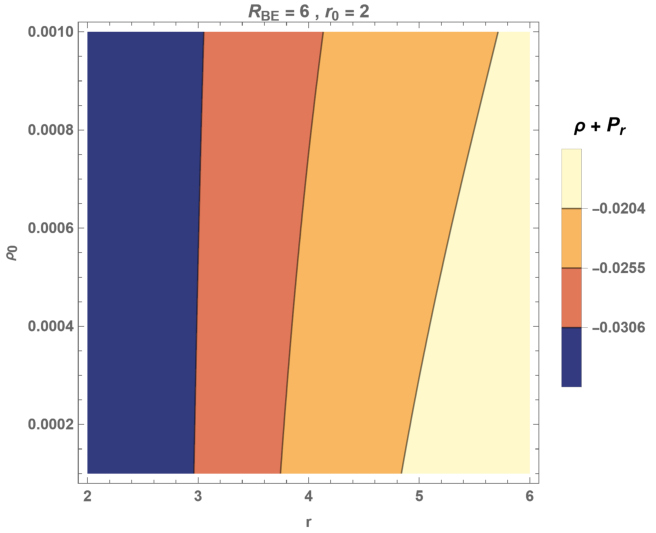} \\
            \includegraphics[width=0.4\linewidth]{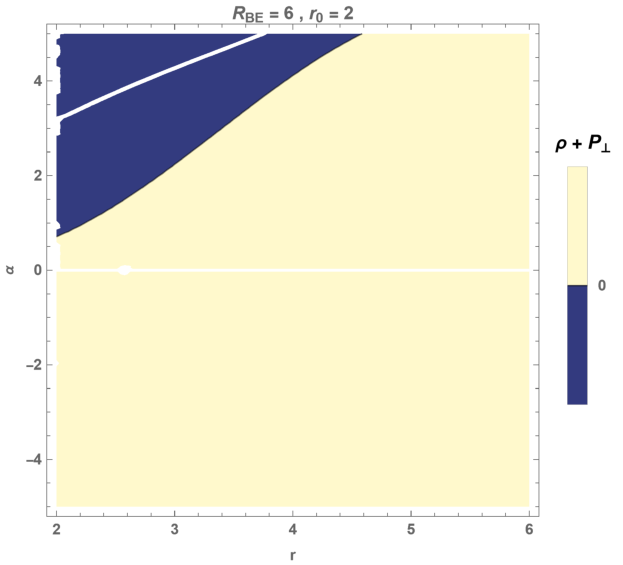}   &
            \includegraphics[width=0.4\linewidth]{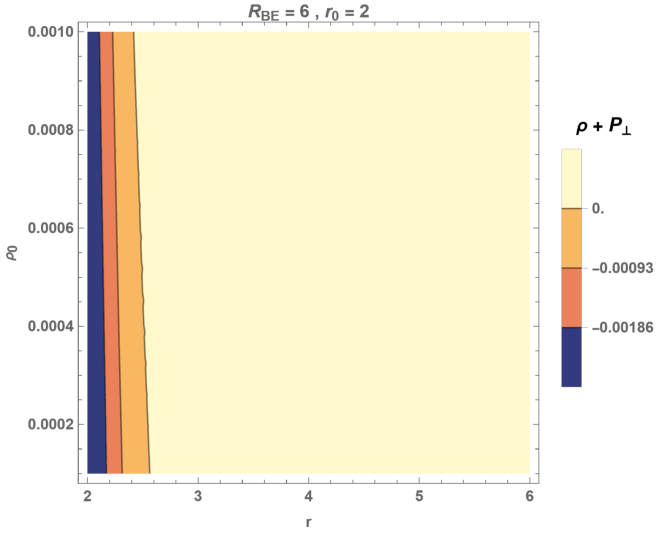}
        \end{tabular}
        \caption{Radial variation of $(\rho(r) + P_r(r))$ and $(\rho(r) + P_{\perp}(r))$ for different model parameters: $(R_{BE},r_0,\rho_0) = (6,2,0.0025)$(top-left and bottom-left plot), \& $(R_{BE},r_0,\alpha) = (6,2,2)$(top-right and bottom-right plot). }        \label{fig:ec2}
    \end{figure}

\begin{figure}[t]
        \centering
        \begin{tabular}{cc}
            \includegraphics[width=0.4\linewidth]{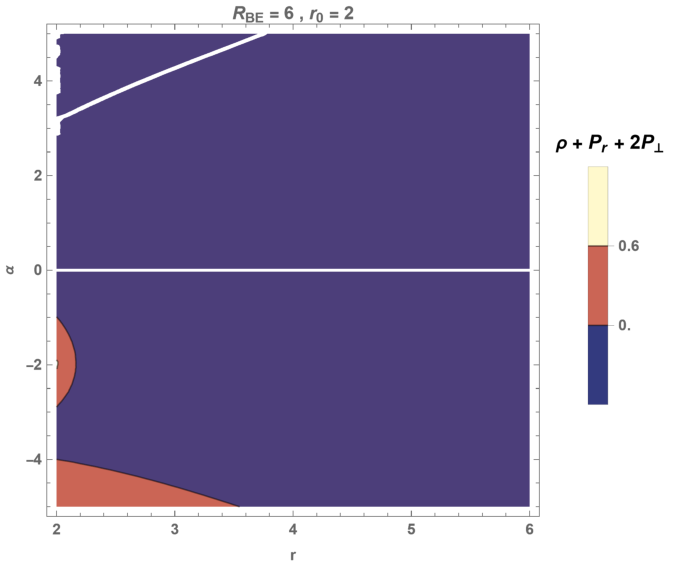}   &
            \includegraphics[width=0.4\linewidth]{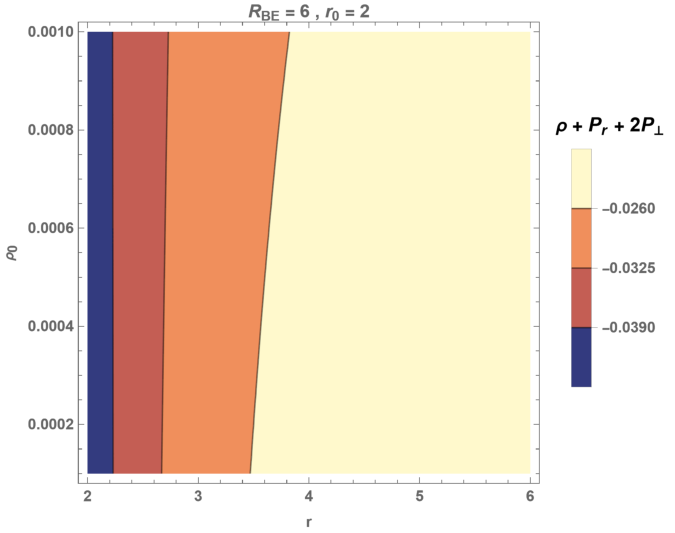} 
        \end{tabular}
        \caption{Radial variation of $(\rho(r) + P_r(r) + 2P_{\perp}(r))$ for different model parameters: $(R_{BE},r_0,\rho_0) = (6,2,0.0025)$(left plot), \& $(R_{BE},r_0,\alpha) = (6,2,2)$(right plot). }
        \label{fig:ec3}
    \end{figure}

\begin{figure}[t]
        \centering
        \begin{tabular}{cc}
            \includegraphics[width=0.4\linewidth]{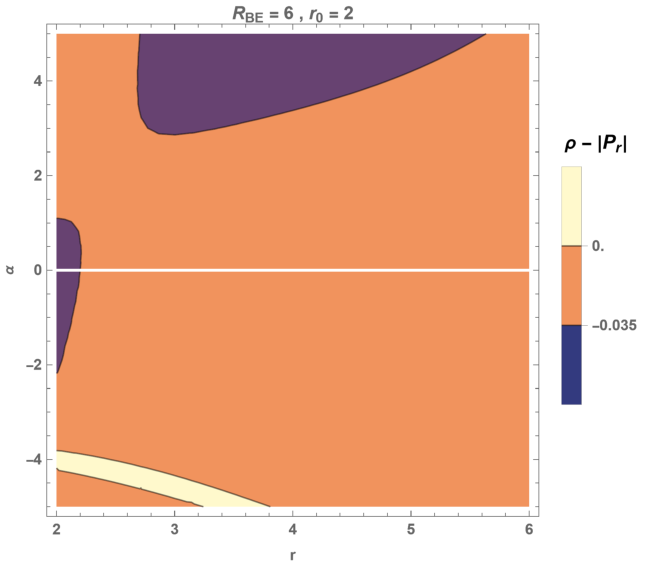}   &
            \includegraphics[width=0.4\linewidth]{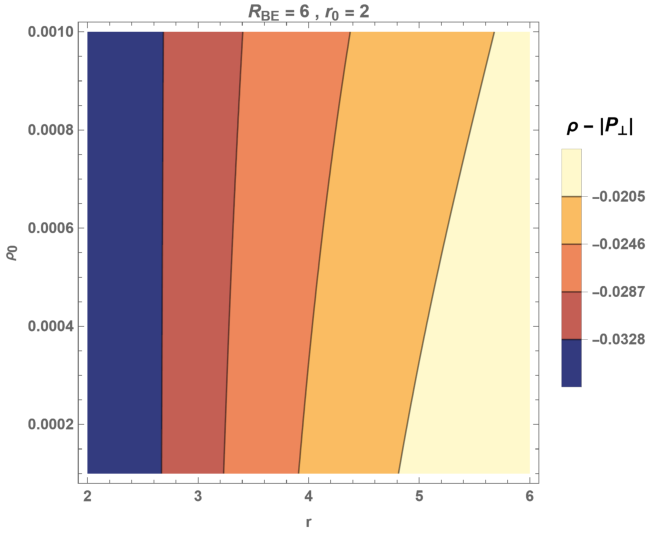} \\
            \includegraphics[width=0.4\linewidth]{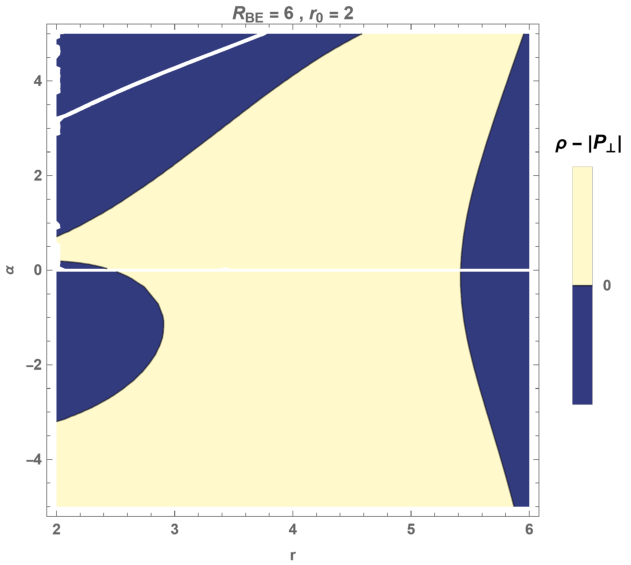}   &
            \includegraphics[width=0.4\linewidth]{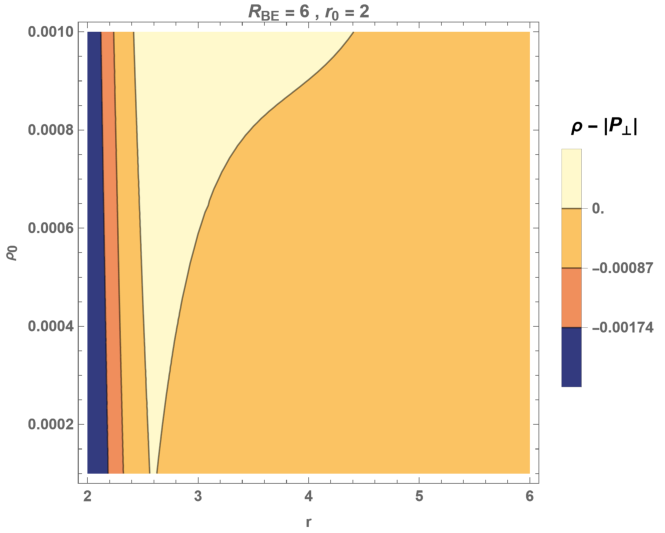}
        \end{tabular}
        \caption{Radial variation of $(\rho(r) - |P_r(r)|)$ and $(\rho(r) - |P_{\perp}(r)|)$ for different model parameters: $(R_{BE},r_0,\rho_0) = (6,2,0.0025)$(top-left and bottom-left plot), \& $(R_{BE},r_0,\alpha) = (6,2,2)$(top-right and bottom-right plot). }
        \label{fig:ec4}
    \end{figure}

The presence of exotic matter near the throat of the WH implies that the null energy condition (NEC) is violated. Violation of the NEC leads to violation of WEC, SEC, and DEC. To test ECs we plot the inequalities in Fig. \ref{fig:ec1} - \ref{fig:ec4} and note different features. The inequalities are examined for a domain of model parameters permitted by a realistic flaring-out condition in Fig. \ref{fig:bprime}. From Fig. \ref{fig:ec1}, it is observed that the density remains positive definite in the WH. It is maximum at the WH throat, which is found to vanish at the BEC radius ($R_{BE}$).

It is evident from figures \ref{fig:ec2} - \ref{fig:ec4} that most of the ECs are violated at the throat of the WH for both positive and negative values of the Gauss-Bonnet coupling parameter. In addition, we found a range of the Gauss-Bonnet coupling parameter for which the EC inequalities are satisfied: $\left( \rho(r) + P_{r}(r) \right) \geq 0$ for $\alpha \in [-3.805,-21.025]$; $\left( \rho(r) + P_{\perp}(r) \right) \geq 0$ for $\alpha \leq 0.71$: $\left( \rho(r) + P_r(r) + 2P_{\perp}(r) \right) \geq 0$ for $\alpha \in [-1,-2.88] \cup [-4,-\infty]$; $\left( \rho(r) - |P_{r}(r)| \right) \geq 0$ for $\alpha \in [-3.805,-4.222]$; and $\left( \rho(r) - |P_{\perp}(r)| \right) \geq 0$ for $\alpha \in [-0.615,-0.195] \cup [-3.2,-\infty]$. This range for the Gauss-Bonnet parameter is obtained for a set of parameters $(R_{BE},r_0,\rho_0) = (6,2,0.0025)$). It is evident from the analysis that there exists a range of the Gauss-Bonnet coupling parameter $\alpha \in [-4,-4.222]$ with  $(R_{BE},r_0,\rho_0) = (6,2,0.0025)$) for which the energy conditions are satisfied at the throat of the WH in the EGB gravity with BEC matter. For other values of the parameters, the energy conditions are found to violate even at the WH throat, which indicates the presence of exotic matter. 

\section{Conclusion}
\label{sec:6}
Wormholes connect two distinct regions of spacetime or two distinct universes. The two distinct regions of spacetime are linked by a non-singular region called a throat, and it was speculated that exotic matter is needed at the throat of the WH to keep the throat open. To realize the above criteria, it is found that an open throat is possible when the null energy condition (NEC) is violated. In the paper, new wormhole solutions is obtained in the 4D Einstein-Gauss-Bonnet (EGB) theory of gravity, incorporating non-relativistic Bose-Einstein condensate (BEC) matter as a phenomenologically motivated profile.

Considering the density profile for BEC matter and a constant redshift function (i.e. $\psi = 1$) we determine the radial and transverse pressures. The shape function ($\mathcal{S}(r)$) of the WH is drawn in Fig. \ref{fig:shapefunc} for a domain of model parameters: $(R_{BE},\,r_0,\,\alpha) \equiv (6,2,1)$ and $\rho_0 \in [0.0001,0.005]$ (top-left); $(R_{BE},\,r_0,\,\rho_0) \equiv (6,2,0.0025)$ and $\alpha \in [-5,5]$ (top-right); $(R_{BE},\,\rho_0,\,\alpha) \equiv (6,0.0025,1)$ and $r_0 \in [0,6]$ (bottom). The asymptotic flatness condition is plotted in Fig. \ref{fig:bbyr} for the above model parameters. We determine the domain of model parameters corresponding to Fig. \ref{fig:bprime} for which the flaring-out condition ($\mathcal{S}'(r) < 1$) is satisfied. The red region in the Fig. indicates that the flaring-out condition is violated.
The proper radial distance, VIQ, and energy conditions are determined for a set of model parameters, which permits an acceptable flaring-out condition. The embedding diagram of the WH and its visualization of the entire WH are also shown in Fig. \ref{fig:WHvisual}. In Fig. \ref{fig:radialdistance}, it is evident  that the proper radial distance decreases from the upper universe as one allows to move towards the throat, which is physically acceptable. Using Fig. \ref{fig:viq} for a set of model parameters, we estimate the exotic matter that stabilizes the WH. We also noted that for $\alpha \leq -4$ the VIQ is greater than zero, indicating that the NEC is not violating for this range of $\alpha$. The anisotropy factor is analyzed in Fig. \ref{fig:anisotropy}, and we found that the anisotropy factor is negative for $\alpha \leq -4$, i.e., it exhibits an inward anisotropy force. We analyze the energy conditions at the WH's throat for a set of model parameters. The radial variation of the energy density in Fig. \ref{fig:ec1} for the following model parameters ($(R_{BE},r_0,\rho_0) = (6,2,0.0025)$, \& $(R_{BE},r_0,\alpha) = (6,2,2)$), is found to admit finite energy density, which is maximum at the throat subsequently vanishes at a distant region corresponding to the BEC radius. In figures \ref{fig:ec2} - \ref{fig:ec4}, we plot the ECs: $\rho + P_r \geq 0$, $\rho + P_{\perp} \geq 0$, $\rho + P_r + 2 P_{\perp} \geq 0$, $\rho - |P_r| \geq 0$, and $\rho - |P_{\perp}| \geq 0$ for the above set of model parameters. It is found that the ECs are satisfied for a given range of the Gauss-Bonnet coupling parameter, which is evaluated: $\left( \rho(r) + P_{r}(r) \right) \geq 0$ for $\alpha \in [-3.805,-21.025]$; $\left( \rho(r) + P_{\perp}(r) \right) \geq 0$ for $\alpha \leq 0.71$: $\left( \rho(r) + P_r(r) + 2P_{\perp}(r) \right) \geq 0$ for $\alpha \in [-1,-2.88] \cup [-4,-\infty]$; $\left( \rho(r) - |P_{r}(r)| \right) \geq 0$ for $\alpha \in [-3.805,-4.222]$; and $\left( \rho(r) - |P_{\perp}(r)| \right) \geq 0$ for $\alpha \in [-0.615,-0.195] \cup [-3.2,-\infty]$. We obtain the range for the Gauss-Bonnet parameter ($\alpha$) with  $(R_{BE},r_0,\rho_0) = (6,2,0.0025)$). It is evident that the energy conditions are satisfied at the throat of the WH for $\alpha \in [-4,-4.222]$ and $(R_{BE},r_0,\rho_0) = (6,2,0.0025)$). It is shown in ref. \cite{Jusufi2020} that the inequality ($\rho + P_r + 2 P_{\perp}$) is satisfied at the throat of an anisotropic WH in the framework of 4D EGB gravity for a small positive value of the Gauss-Bonnet coupling parameter, i.e., $\alpha = 0.1$. Also the inequality ($\rho + P_{\perp}$) is satisfied for $\alpha = 0.1$ at the throat of a WH depending on the density function $\rho = \rho_0 \left( r_0/r \right)^{\beta}$ in the framework of 4D EGB gravity. However, all the cases discussed in the paper violate the classical energy conditions in general for $\alpha > 0$. In the present work, we obtain a range of the Gauss-Bonnet parameter ($\alpha$) for which the energy conditions are satisfied in the 4D EGB gravity with BEC matter, which is a new result. The stability analysis in Fig. \ref{fig:stability} shows that the WH model attains vanishing sound speed ($C_s^2 = 0$) at the throat for the Gauss-Bonnet coupling parameter value $\alpha = -0.0512471$ (for model parameters: $(R_{BE},r_0,\rho_0) = (6,2,0.0025)$). We also note that the WH model is causal ($0 \leq C_s^2 < 1$) within the limit $\alpha \in [-0.0512471,-0.0770328]$. We obtain WH solutions in 4D EGB gravity with BEC matter for both positive and negative Gauss-Bonnet coupling parameter. It is interesting to note that the WH solutions are permissible for $\alpha <0$ which is different from the string theory perspective. In the past a realistic cosmological model was obtained in higher dimensions \cite{Paul1990} with negative alpha, our result supports that found in the paper. We also note that  for a given negative value of $\alpha$ WH are permitted satisfying all the energy conditions.

We conclude that non-relativistic BEC matter supports the formation of a stable traversable WH in 4D EGB gravity. Although wormholes have not been detected yet, we have investigated the possible existence of traversable WHs in this study in the 4D EGB gravity with BEC dark matter. In this paper, we have chosen a constant redshift function, however, it will be interesting to explore WH geometries with non-constant redshift functions in future.

\begin{acknowledgements}
BD is thankful to CSIR, New Delhi, for financial support. BCP acknowledges support from SERB, Govt. of India (F.No. CRG/2021/000183) for awarding a project. The authors are thankful to ICARD, NBU, for extending research facilities.
\end{acknowledgements}

\bibliographystyle{spphys}       
\bibliography{referencesWH}   


\end{document}